\documentclass{aa}  

\usepackage{graphicx}
\usepackage{txfonts}
\begin{document} 

\title{Three dimensional MHD Modeling of Vertical Kink Oscillations in an Active Region Plasma Curtain}
\titlerunning{3D MHD Modeling of Vertical Kink Oscillations in a Plasma Curtain}

\author{L. Ofman\inst{1,2,3}
  \and  M. Parisi\inst{4}
 \and  A. K. Srivastava\inst{5}}

\institute{Catholic University of America, Washington, DC
20064
\email{Leon.Ofman@nasa.gov}
\and NASA Goddard Space Flight Center, Code 671,
Greenbelt, MD 20771 
\and Visiting, Department of
Geophysics and Planetary Sciences, Tel Aviv University, Tel Aviv 69978,
Israel
\and Department of Mechanical and Aerospace Engineering, Sapienza University of Rome, via Eudossiana, 18, 00184, Rome, Italy
\and Department of Physics, Indian Institute of Technology (Banaras Hindu University), Varanasi-221005, India}

\date{}

\abstract
{Observations on 2011 August 9 of an X~6.9-class flare in active region (AR) 11263 by the Atmospheric Imaging Assembly (AIA) on-board the Solar Dynamics Observatory (SDO), were followed by a rare detection of vertical kink oscillations in a large-scale coronal active region plasma curtain in EUV coronal lines. The damped oscillations with periods in the range 8.8-14.9 min were detected and analyzed recently.}
{Our aim is to study the generation and propagation of the MHD oscillations in the plasma curtain taking into account realistic 3D magnetic and density structure of the curtain. We also aim at testing and improving coronal seismology for more accurate determination of the magnetic field than with standard method.}
{We use the observed morphological and dynamical conditions, as well as plasma properties of the coronal curtain based on Differential Emission Measure (DEM) analysis to initialize a 3D MHD model of its vertical and transverse oscillations by implementing the impulsively excited velocity pulse mimicking the flare generated nonlinear fast magnetosonic propagating disturbance interacting with the curtain obliquely. The model is simplified by utilizing initial dipole magnetic field, isothermal energy equation, and gravitationally stratified density guided by observational parameters.}
 {Using the 3D MHD model, we are able to reproduce the details of the vertical oscillations and  study the process of their excitation by nonlinear fast magnetosonic pulse, propagation, and damping, finding agreement with the observations.} 
{We estimate the accuracy of simplified slab-based coronal seismology by comparing the determined magnetic field strength to actual values from the 3D MHD modeling results and demonstrate the importance of taking into account more realistic magnetic geometry and density for improving coronal seismology. }

\maketitle

\section{Introduction}
With the launch of the Solar Dynamics Observatory (SDO) the detection of waves in Extreme Ultraviolet (EUV) emission in the corona using Atmospheric Imaging Assembly (AIA) instrument became possible in many observed events with potential applications to coronal seismology (CS) to understand the physical conditions in the corona. Using SDO/AIA EUV channels transverse coronal loop oscillations were observed to be ubiquitous in active region loops \citep{Mci11} and were analyzed in great detail with application to coronal seismology \citep[e.g.,][]{AS11,YN12,WV12,Wan12,Nis13,Ver13a,Ver13b,Nis14}. Nevertheless, vertical oscillations in coronal loops or magnetic arcades are rarely observed in the EUV emissions in-spite of high-resolution and high-cadence recent observations by AIA, and past high-resolution observations by the Transition Region and Coronal Explorer (TRACE) satellite. The vertical kink oscillations of a curved coronal loop are fundamentally different from the horizontal kink oscillations, since they are confined to the loop curvature plane, and can lead to the change of loop length. While horizontal kink oscillations were first detected by \citet{Asc99} and \citet{Nak99}, and were studied extensively since then in many papers motivated by their potential use in MHD CS  for localized coronal structures\citep[e.g., reviews  by][and references therein]{NV05,And09,DN12,LO14}, only later \citet{WS04} have reported the first confirmed detection of vertical oscillations in a coronal loop using TRACE 195~\AA~observations of a coronal loop on the limb on 17 April 2002, with additional cases analyzed in \citet{WSS08}. Vertical oscillations in a hot coronal loop following a flare/CME was reported for the first time by \citet{WVF12} using SDO/AIA  131~\AA~($\sim$10 MK) and 94~\AA~($\sim$6.3 MK) EUV bandpasses. Vertical oscillations in cool coronal loops were also detected with SDO/AIA in conjunction with STEREO/EUVI. These oscillating loops were triangulated using STEREO/EUVI data, and oscillation properties derived from SDO/AIA and STEREO/EUVI were utilized for refined coronal seismology \citep{AS11}. Recently, \citet{SG13} (hereafter, SG13) have reported vertical oscillations in a coronal active region plasma curtain observed by SDO/AIA in cool coronal line of 171~\AA~($\sim$1 MK). Vertical transverse oscillations were also observed recently in a coronal magnetic flux rope using SDO/AIA \citep{Kim14}.

While several 2D numerical and analytical models of vertical kink oscillations in the coronal structures were published \citep[e.g.][]{Sel05,Sel06,Ver06,Gru06,Sel07}, the full three dimensional MHD modeling is required to study the complete interaction between the excitation mechanisms and the 3D mode polarizations of the oscillations in curved magnetic field geometry and the impulsively excited propagating nonlinear fast magnetosonic waves launched by a  flare that induces the oscillations. The 3D MHD modeling of a bipolar active region (AR) loop oscillations provides further insights on the physical scenario and inherent cause for the somewhat rare detection of this phenomena \citep{SSO11}. At present, the application of coronal seismology is often based on linear dispersion relation for waves in cylindrical slabs or otherwise simplified structures that compromises the accuracy of the method \citep[e.g.][]{Ofm09,PD14}. In order to improve CS one needs to apply more realistic modeling of the oscillating structures. More realistic, bipolar magnetic field \citep[e.g.][]{Miy04,MO08,SO09,SOS11,SSO11,Sel13,PD14} or extrapolated magnetogram based magnetic field \citep[e.g.,][]{Ofm07,SO10} and gravitationally stratified density structures were modeled with 3D MHD in order to study waves, clearly demonstrating the advantage of considering 3D realistic magnetic field topology on unambiguous determination of seismologically derived coronal magnetic field.

In the present article, we utilize recent observations by SG13 of vertical oscillations of a coronal curtain to setup the parameters of the 3D MHD model and study for the first time the excitation, propagation, and damping of the vertical oscillations in a coronal bipolar curtain with applications and quantitative tests of coronal seismology by comparing to 3D MHD modeling results. The paper is organized as follows: in Section~\ref{obs:sec}, we present the brief summary of the observations, the numerical model is described in Section~\ref{model:sec}, the numerical results are in Section~\ref{num:sec}, and Section~\ref{con:sec} is devoted to the Discussion and Conclusions.

\section{Summary of the Observations and DEM analysis}
\label{obs:sec}
\begin{figure*}
       \includegraphics[width=18cm]{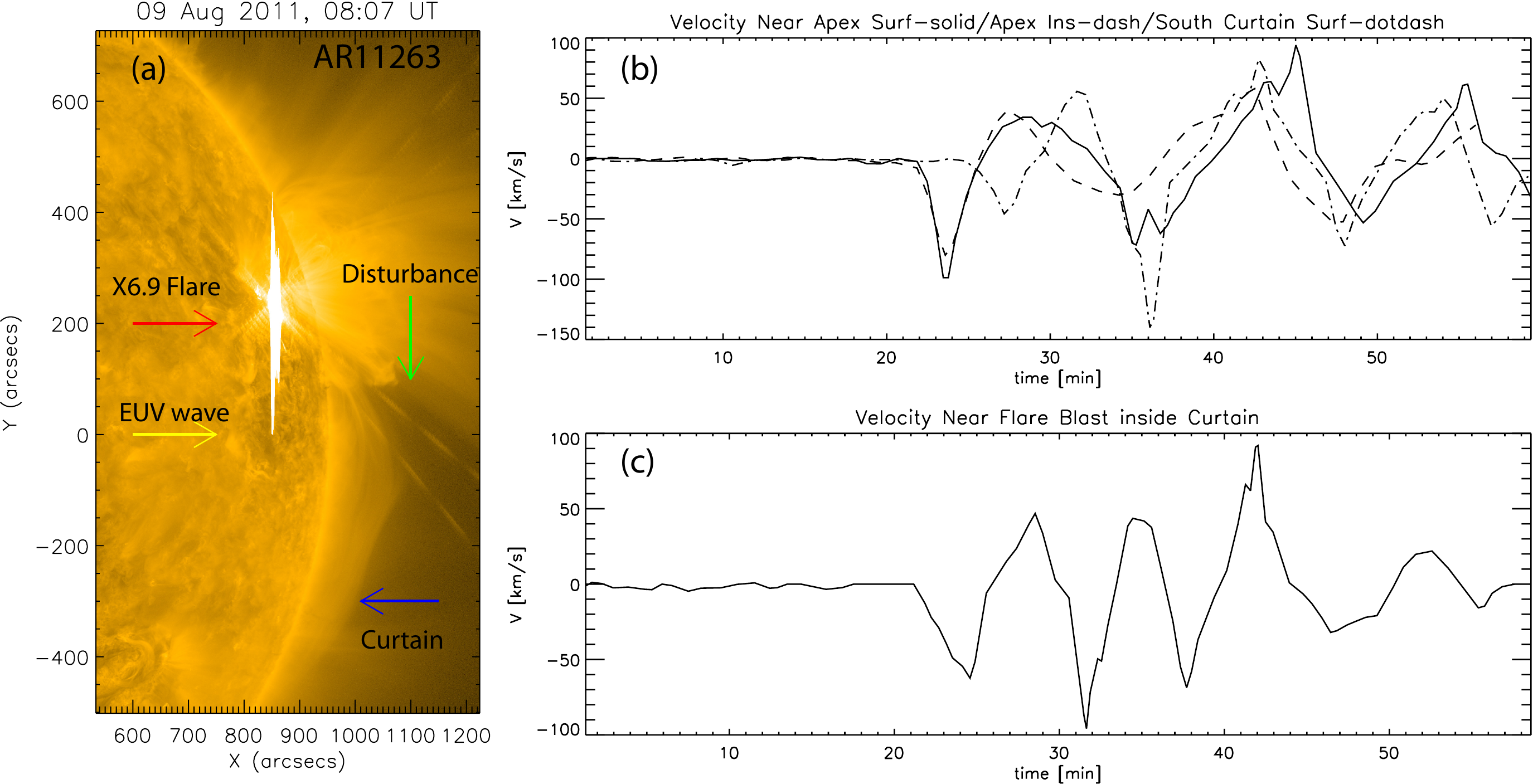}
      \caption{(a) The snapshot on 9 August 2011 at 08:07 displays
the X~6.9 flare (red arrow), EUV wave (yellow arrow), plasma curtain (blue arrow), and
disturbances passing in a narrow channel above it (green arrow) (adapted from SG13). The temporal evolution of is available as a movie in the online edition. (b) The temporal evolution of the velocity in various parts of the plasma curtain as observed
by SG13. (c) The temporal evolution of the velocity near the flare site.
              }
         \label{SG13:fig}
   \end{figure*}

Recently, the detection of a diffused and laminar plasma curtain at western equatorial 
corona that oscillates vertically during passage of the coronal disturbances (EUV waves) triggered by
X~6.9 class solar flare (start: 07:48 UT; peak: 08:05 UT; end: 08:08 UT) in AR 11263 (N17 W69) in its vicinity on 9 August 2011 was reported and analyzed by SG13 (cf., Fig.~\ref{SG13:fig}a). The vertical
kink oscillations are triggered near the flare energy release site within the curtain (8.8 min); in the middle both at the apex (14.9 min) and inside (13.3 min), as well as at the surface in the south-most part of the curtain (12.7 min).   The properties of these excited waves
strongly depend on the local plasma and magnetic field conditions of the plasma curtain. The right-panels of Fig.~\ref{SG13:fig} show the temporal evolution of the velocity in the middle at the surface (Fig.~\ref{SG13:fig}b: solid line), inner part (Fig.~\ref{SG13:fig}b: dashed line), as well as at the surface in the southward part (Fig.~\ref{SG13:fig}b, dotted-dashed line) of the curtain, and near the flare blast site (Fig.~\ref{SG13:fig}c).
These velocity oscillations in the plane of the sky are completely localized in the space at various parts of the curtain and derived here from the time-distance measurement data of SG13. The animation of the oscillation observed with SDO/AIA 171\AA\ channel are enclosed online with the article.
We carried out emission measure and temperature distribution analysis of the western equatorial corona
and plasma curtain during 8-9 UT on 9 August 2011 using the automated method developed by \citet{AscB11}. Estimation of these plasma properties provided clues on the
localized thermal conditions of plasma curtain. We use the full-disk SDO/AIA in all EUV channels at three times covering
the whole oscillation period of the curtain, i.e., 08:07 UT, 08:28 UT, 08:44 UT on 09 August 2011. We calibrate and clean
the data using the {\tt aia\_prep.pro} subroutine of SSWIDL and co-align
the AIA images as observed in various AIA filters by using the
co-alignment test as described by \citet{AscB11}.
Six AIA EUV filter full-disk and co-aligned images (304 \AA, 171 \AA,
193 \AA, 94 \AA, 335 \AA, and 211 \AA) covering temperature in the 
0.5-9.0 MK range, have been utilized to calculate the DEM and produce the temperature maps.
The $\chi^2$ lies mostly below 1.0 in the most of the parts of curtain except its 
top laminar layer and background corona above the curtain. However, in the given limit of the 
uncertainties, within the curtain we have good DEM 
forward fit, and with the least error of estimation the measure of temperatures and emission measures in the region of the EUV curtain. 

\begin{figure*}
       \includegraphics[width=18cm]{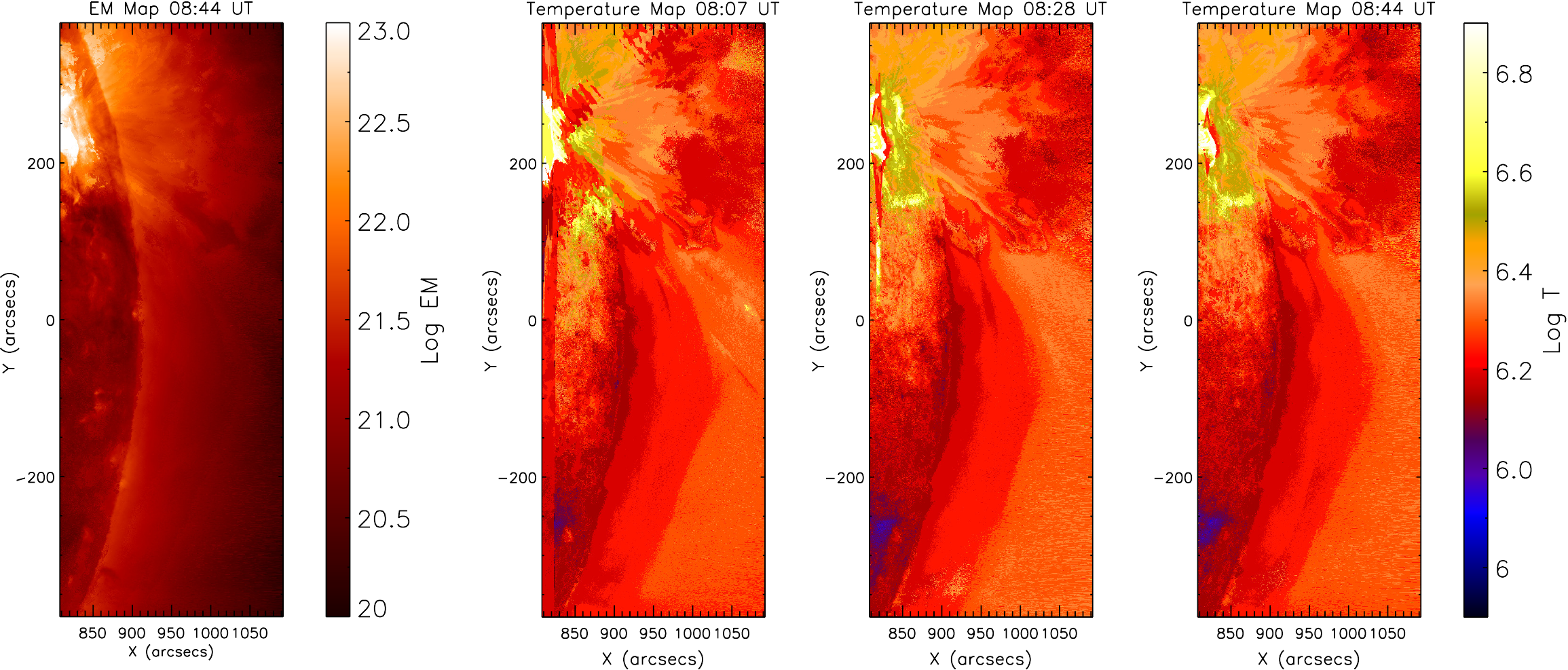}
      \caption{Temperatures maps obtained from the DEM analysis of the coronal curtain shown in Fig.~\ref{SG13:fig} using six AIA EUV lines at times 08:07 UT, 08:28 UT, 08:44 UT on 09 August 2011 (right three panels) The emission measure map at 08:44 UT on 09 August 2011 is shown in the left panel.}
         \label{DEM:fig}
   \end{figure*}

Fig.~\ref{DEM:fig} shows that the plasma curtain (and its oscillatory parts) are nearly at
isothermal temperature at the beginning of the oscillations (8:00 UT), and without significant temperature variability throughout the oscillation sequence. The temperature inside the curtain varies by about 15\% - close to the observed uncertainty. It is evident that hotter material is located above the curtain - and this may explain the darkening of the 171\AA\ emission in this region seen in Fig.~\ref{SG13:fig}, since the peak response of this channel is $\sim$1MK \citep{Lem12}. The hot material in the line-of-sight may also lead to apparently increased DEM temperature inside the curtain. Inspecting the Emission Measure (EM) map proportional to the column density in the line of sight in Fig.~\ref{DEM:fig} (left panel) shows that the height-dependence of the density is consistent with exponentially decreasing density with height. This observation justifies our isothermal and initially hydrostatic approximation model of the plasma curtain that is valid to understand the evolution of the wave 
phenomena in the present study.

\section{The 3D MHD Numerical Model}
\label{model:sec}
For the present study we adapt the numerical model of a bipolar AR initially developed by \citet{OT02}, and used extensively to study magnetosonic waves in AR loops \citep[e.g.][]{MO08,SO09,SOS11,SSO11,OWD12,WOD13}. Note, it is not the goal of the present study to model the detailed magnetic, density, or thermal structure of the curtain. By using this model we aim to improve the MHD seismology of the observed vertical kink oscillations using more realistic field and density structure than used in CS based on wave dispersion from simplified slab geometry. The improvement is achieved by matching the parameters of the global bipolar structure of the curtains' field and the gravitational density stratification {to the observationally deduced parameters. The parameters of the model were adjusted such that the idealized model is in good agreement with the dynamics of oscillating laminar and diffused large-scale plasma curtain due to the X~6.9 class flare generated disturbances reported by SG13. In particular, the aspect ratio of the dipole was selected to fit the observed active region curtain aspect ratio, the temperature that determines the gravitational scale height of the density was selected to match the observed line emission temperature, the normalizing Alfv\'en speed and the parameters of the velocity Gaussian pulse launched at the boundary were optimized through parametric study to match the observed wave amplitudes -  further details given below.  

In order to model the event, we solve numerically the time-dependent, nonlinear, finite-$\beta$, resistive 3D MHD equations with gravity, using a Cartesian grid in the range $\left( -3.5, 3.5 \right) \times \left( -3.5, 3.5 \right) \times \left( 1.0, 6.0\right)$, in normalized units (the normalizations are given below). 
The adopted method for integrating the partial differential equations is a modified Lax-Wendroff method with fourth-order stabilization term \citep[see,][]{OT02}. Furthermore, we use \citet{Pow94} method to correct for any numerical nonzero divergence of the magnetic field. The set of equations includes the contribution from gravity and the energy equation is eliminated by the isothermal approximation (polytropic index 
$\gamma = 1$). While using $\gamma=5/3$ would be appropriate for thermodynamic equilibrium
conditions not applicable in the corona, or when the coronal heating terms and losses are known in detail and   computed in the energy equation, the polytropic (isothermal) approximation is based on the reasonable approximation that the heating and losses terms are balanced exactly in the energy equation, and combined with the high coronal heat conduction lead to constant temperature. The nearly isothermal structure of coronal loops in quiescent active regions has been deduced from past observations \citep[e.g.,][]{Asc05}.

 Moreover, observational evidence of coronal loop oscillations shows that the effective polytropic index is close to unity \citep{Van11}, in line with our choice of polytropic index. This approximation is supported by the DEM analysis of the present structure showing about 15\% variability of the temperature in the active region curtain in Fig.~\ref{DEM:fig}, on the order of the observational uncertainty. Since in the oscillation in the curtain are observed for about 40 min, the  effects of resistive dissipation can be neglected. At the end of the oscillations the intensity structure of the curtain reverts mostly to the original form, indicating negligible net effect of the oscillations on the curtain. Non-isothermal (cooling) loop oscillations were studies in the past  \citep[e.g.,][]{ME09,Erd11,Rud11,Alg14}. However, in the present study we neglect the effect of cooling, since the observed active region temperature remains nearly constants on the oscillation timescale, as evident in Fig.~\ref{DEM:fig}. 

 Multiple processors (up to 144) are used to execute the code, which is parallelized using Message Passing Interface (MPI). We assume that the magnetic field perturbed by the flare can be reasonably described by perturbing the potential dipole (Fig.~\ref{3Dinit:fig}a) in the 3D computational domain, modeled on a 512$\times$512$\times$434 grid. The high resolution resolves well the background density and magnetic field gradients as well as the propagating fast magnetosonic pulse launched to model the effects of the flare, and minimizes the possible effect of numerical dissipation - this was tested by comparing to lower resolution runs. The foot-points are located at ($\pm$2.5, 0.0, 1.0) in the box of normalized units, such geometry allows to preserve the aspect ratio of the observed plasma curtain ($\sim$ 530 Mm length and height of $\sim$100 Mm).

The MHD equations are normalized using the following parameters: length scale of $a = R_s / 10$, magnetic field at the foot-points $B_0\,=\,23\,G$, temperature $T_0\,=\,1\,MK$ is the isothermal temperature in the model active region curtain, particle density $n_0\,=\,10^9\,cm^{-3}$ at $z_{min}=1$. These values are guided by the observationally determined values obtained by SG13 and determine the Alfv\'en speed of $V_{A0}\,=\,1586.5\,km/s$, the  Alfv\'en time $\tau_A\,=\,44.1\,s$, and the isothermal sound speed of $C_s\,=\,\left(p_0 / \rho_0 \right)^{1/2}\,=\,129\,km/s$.

The observed oscillations shown in Fig.~\ref{SG13:fig} and modeled here are for plasma emitting from Fe~IX at 171\AA\ in EUV. Since the peak response of the 171\AA\ EUV filter is around 1MK \citep{Lem12} the observed oscillating magnetic structure shows 1MK plasma component in the multithermal coronal plasma, and justifying the choice of temperature for the numerical model of the present observation. Further, as discussed above it is evident from the DEM analysis in Fig.~\ref{DEM:fig} the DEM temperature is nearly constant inside the active region curtain during the observed oscillation period. 
\begin{figure*}[ht]
	\centering
        \includegraphics[width=9cm]{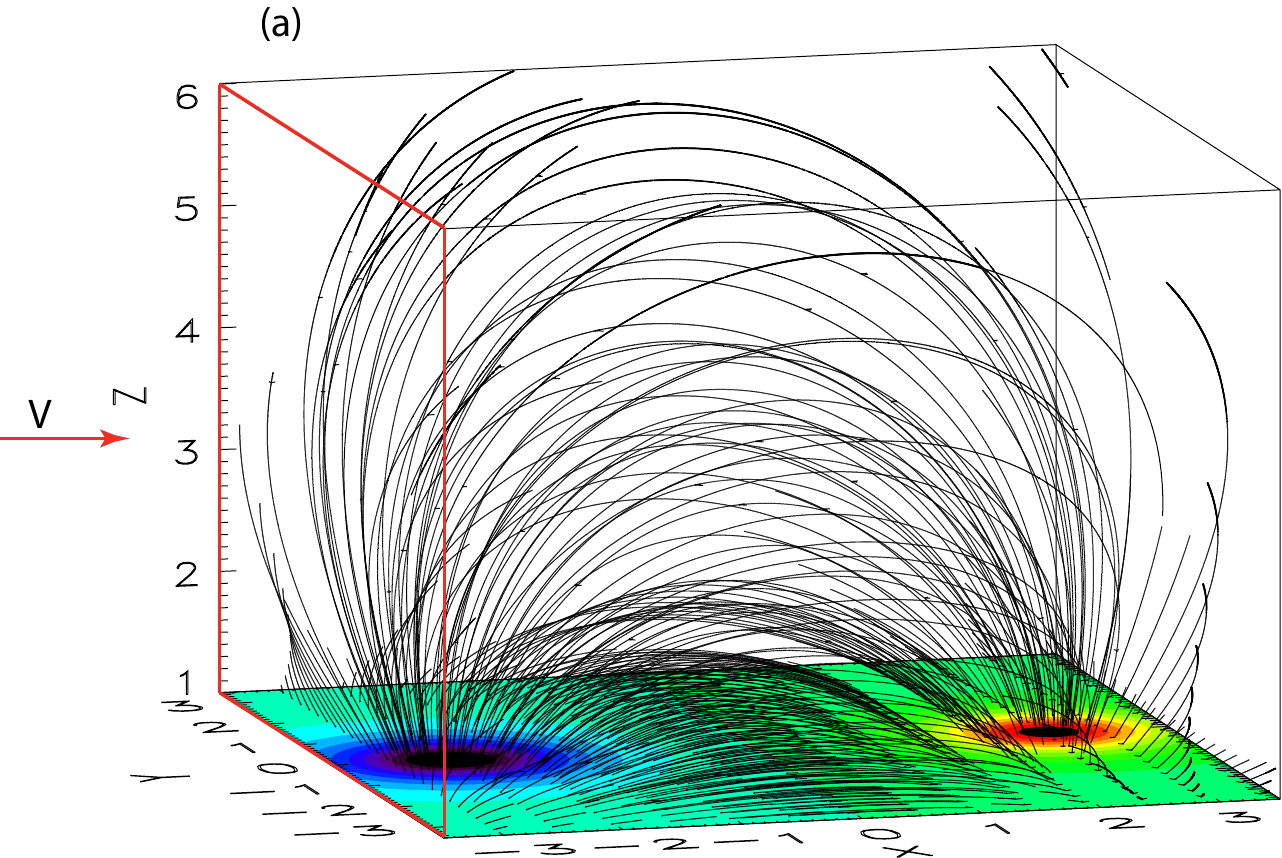}
        \includegraphics[width=14cm]{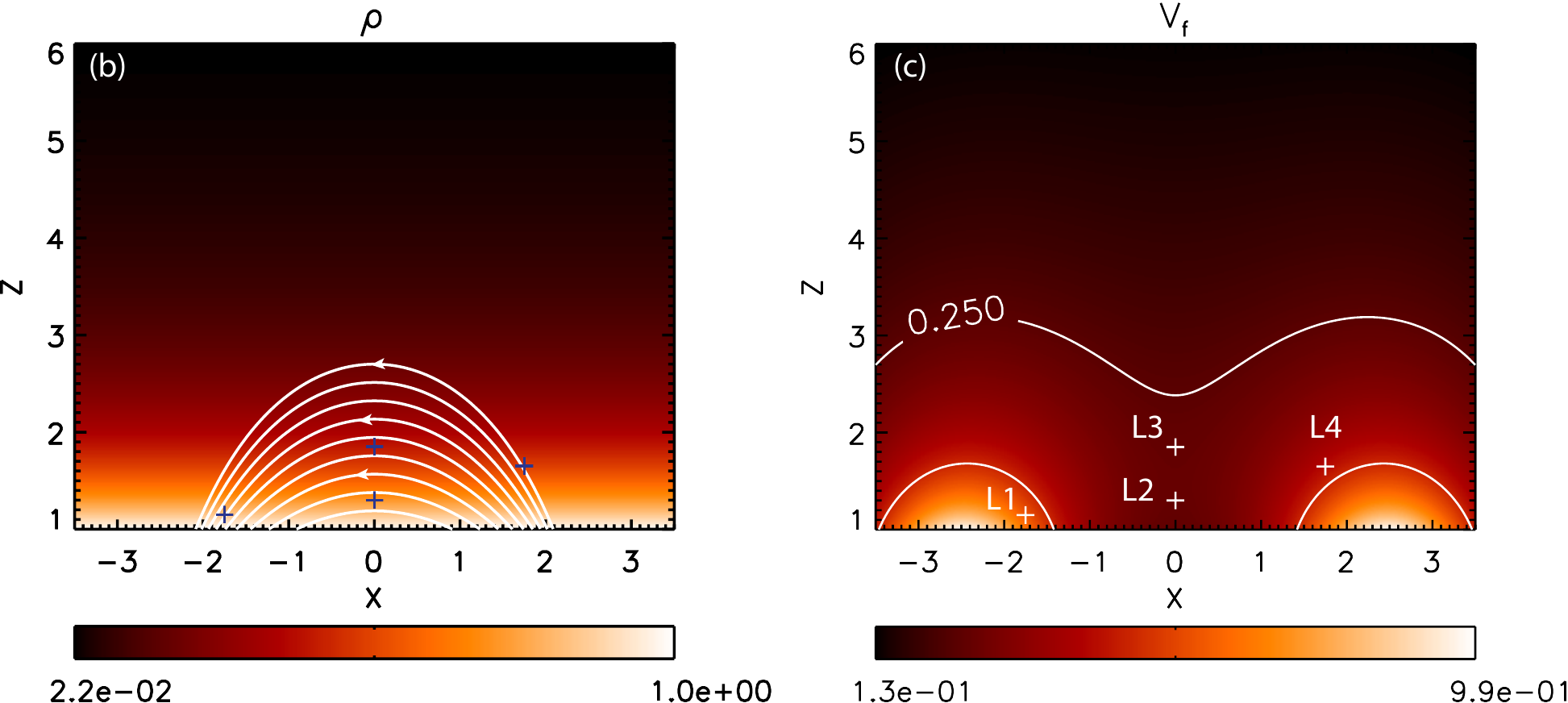}
 	\caption{\textit{(a)} Idealized model of the AR as a magnetic dipole. The red box shows the $y_z$ boundary plane at $x=0$ impacted by the propagating disturbance that models the effect of the flare on the active region curtain. The red arrow shows the direction of the impacting velocity pulse applied as the time-dependent boundary condition on this plane. The orientation of the model is turned by 90$^o$ CCW with respect to the observation shown in Fig.~\ref{SG13:fig}. Bottom: Normalized initial density distribution overlayed with some of the closed field lines that belong to the curtain are shown. \textit{(b)} and initial fast magnetosonic speed \textit{(c)} in the \textit{xz} plane at y=0. The blue (white) `+' symbols mark the locations of the points $L1-L4$ used for temporal evolution diagnostic (see below).}
 	\label{3Dinit:fig}
\end{figure*}
The selected value for the Alfv\'en speed exceeds the value estimated by SG13 using simplified (slab) coronal seismology by about 20\%. We found, in fact, that we need to adopt a larger value by increasing the background magnetic field $B_0$, in order to match the observed oscillation amplitude. One must not forget that in our work we are modeling MHD oscillations of a complex curved magnetic structure such as a coronal  curtain by using an idealized model for the AR based on a dipole magnetic field, thus it is expected that the model values of the physical parameters improve on a simplified slab model, but may still differ from the actual values. 

The initial density profile is assumed to be gravitationally stratified and characterized by a normalized scale height $H$
\begin{eqnarray}
	\rho=\rho_0 e^{[1/(10+z-z_{min})-1/10]/H},
	\label{rho:eq}
\end{eqnarray} where $H=2 k_B T_0 R_s / \left( 10 G M_s m_p \right)$, $R_s$ is the solar radius, $k_B$ is Boltzmann constant, $T_0$ is the temperature, $G$ is the universal gravitational constant, $M_s$ is the solar mass and $m_p$ is the proton mass. The initial density distribution is showed in Fig.~\ref{3Dinit:fig}a, as a cut in the \textit{xz} plane ($y=0$), where \textit{x} is the direction of the initial injection of the idealized disturbance that models the effects of the flare (see below). Some closed magnetic field lines that belong to the curtain in this plane are shown.  Although the image in Fig.~\ref{SG13:fig}a shows higher emission in the curtain than in the corona above this is likely due to the temperature increase of the ambient plasma  above the curtain as evident from the DEM analysis in Figure~\ref{DEM:fig}. The narrow band 171\AA\ AIA channel is sensitive to temperatures around 1MK \citep{Lem12} and the hotter material evident above the curtain appears darker in this emission. However, inspecting the emission measure map of this region in Figure~\ref{DEM:fig} (left panel) does not show sharp transition boundary and is consistent with exponentially decreasing density used in our model. In our model the geometry of the curtain is modeled by the dipole field and the (nearly) frozen-in plasma carried by the fieldlines during the oscillations. Moreover, our analysis is focused on the regions inside the curtain  with observationally guided height obtained from the magnetic loops aspect ratio and the possible small effect on the transverse waves in the interface region between the cooler and hotter region is not considered.

In Fig.~\ref{3Dinit:fig}b, we show the normalized fast magnetosonic speed (in terms of $V_{A0}$), $V_f \left( x, z \right) = \left[{V_A \left( x, z \right)^2+ C_s^2}\right]^{1/2}$ for perpendicular propagation in the \textit{xz} plane for $y=0$, of the initial state. As it is clear from the image, $V_f$ speed decreases extremely fast with the distance from the footpoints, reducing to 25\% of the maximum value at about just half the height of the grid ($z = 3.5$). This is because, for $C_s \ll V_A$, as we can reasonably assume in this case, $V_f$ behaves basically as $V_A$, thus decreasing as $B \sim 1/r^3$ (if far enough from the dipole) and being inversely proportional to $\rho^{1/2}$, which, in turn, decreases nearly exponentially with $z$ and is characterized by a scale height $H$ that depends upon the observationally guided choice of the initial temperature of the coronal plasma in the curtain. The location of the points $L1-L4$ used for temporal evolution diagnostic are marked with '+' symbols in the \textit{xz} plane, where the 3D coordinates in normalized units of the points are $L1=(-1.75,-3.35,1.15)$, $L2=(0.0,1.7,1.3)$, $L3=(0.0,1.7,1.85)$, $L4=(1.75,-0.9,1.65)$. The choice of these points is consistent with the observing points in SG13 shown in Fig.~\ref{SG13:fig}.

Boundary conditions are imposed on each surface at the edges of the 3D computational box. Open boundaries with zero-order extrapolation for every plasma variable are placed at all boundaries except for the lower boundary ($z=z_{min}$), where we impose a condition of wave reflection simulating line-tied boundary, and for the interface between the integration domain and the external disturbances. Since modeling the flare is beyond the scope of the present study in order to trigger the vertical kink oscillations in the plasma curtain as observed by SG13, we perturb the AR initial state with a Gaussian shaped velocity pulse applied as time dependent boundary condition for short duration compared to typical normal mode periods (similar in form to the ones used in \citet{SO09,SOS11}) in the \textit{x} direction (see Figure~\ref{3Dinit:fig}a)
\begin{eqnarray}
	V_x = A_V exp \left[- \left( \frac{ z - z_{min}}{w_z} \right)^2 \right]exp \left[- \left( \frac{t - \left( t_1 + 2 \delta t \right)}{\delta t} \right)^2 \right],
	\label{eq:pulse}
\end{eqnarray} applied at the $yz$ plane at $x=x_{min}$ for the duration of the pulse $\delta t = t_2 - t_1 = 2 \tau_A$}. The pulse magnitude is normalized in units of $V_{A0}$, and  the parameters are $A_V = 0.18$, $w_z = 0.7$, where $t_1$ is the pulse start time at $0.5 \tau_A$ after the model initiation time $t=0$. The value of $w_z$ provides the width of the Gaussian pulse in $z$ and deposits the energy in the lower part of the active region curtain, consistent with observations. The values of the pulse parameters have been chosen by performing a parametric study that best fits the observed wave amplitude at point L1-L2. For example, we have repeated the same model runs with several valued of $A_V$ in the range $A_V=0.05$ to $A_V=0.25$, and found through interval halving of this parameter that $A_V=0.18$ provides oscillation amplitudes that agree well with observations. However, the periodicity and the damping rate of the waves are not sensitive to the choice of the pulse parameters for small $\delta t$ compared to the periods. The value of $B_0$ were varied in several runs in the range 19-26 G, and as expected, it was found that the oscillation period is proportional to $B_0$ with other parameters fixed, providing best fit for $B_0=23$ G.  

\section{Numerical Results}
\label{num:sec}
In this section we highlight the  results of our MHD model of flare-induced vertical kink oscillations excited by means of a lateral Gaussian-shaped pulse, modeling the effects of the kinetic energy release by the X~6.9-class solar flare. 
\begin{figure*}[ht]
	\centering
   	{\includegraphics[width=16cm]{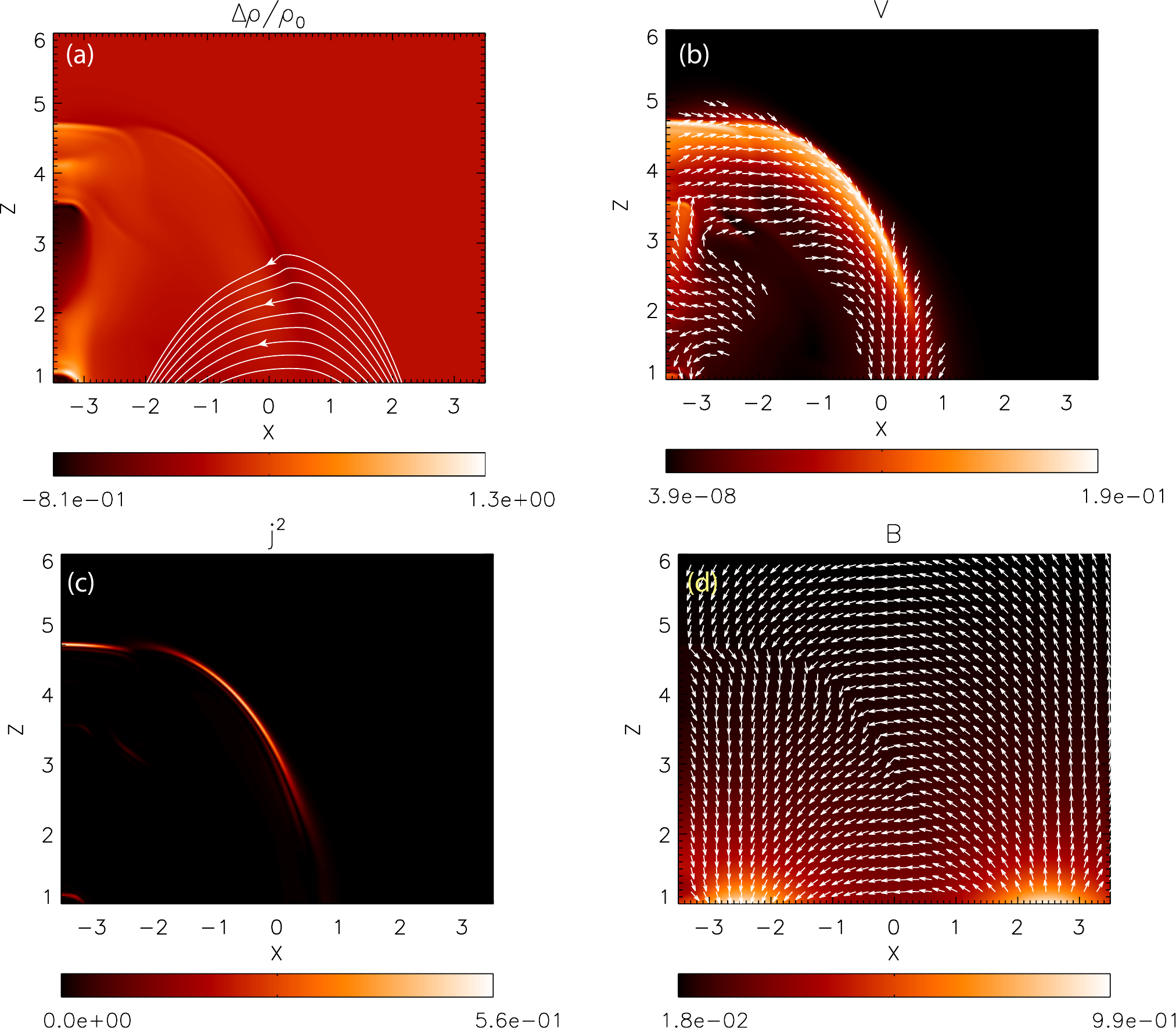}}
 	\caption{Snapshot of the cuts in the \textit{xz} plane at the center of the curtain obtained with the 3D MHD model at $t=14.7 \tau_A$ showing the propagating fast magnetosonic disturbance produced by the boundary pulse that models the effects of the flare. (a) The perturbed density $\Delta \rho/\rho(t=0)$. Fieldlines associated with the active region curtain are shown.  The distortion of the field lines by the propagating fast magnetosonic disturbance is evident. (b) The velocity magnitude and the direction (arrows are shown for $|v|$ greater than 6\% of the absolute maximal velocity). (c) The current density map $j^2$. (d) The magnitude of the magnetic field and the direction. The results show the effects of the initial induced wave pulse and the response of the curtain. 
}
 	\label{vcomp:fig}
\end{figure*}

Fig.~\ref{vcomp:fig} shows the evolved perturbation in terms of plasma velocity, perturbed plasma density (defined as $\Delta\rho=[\rho(t)-\rho(0)]/\rho(0)$, and the magnetic field in the \textit{xz} plane. The snapshot has been captured at $t=14.7\tau_A$, at $y=0$, where $x$ is the injection direction of the velocity pulse. The form of the velocity pulse and the associated compression of the density are shown in Fig.~\ref{vcomp:fig}a-b. The current density associated with the fast magnetosinic wave front produced by the pulse, and the corresponding changes of the magnetic field direction are evident in Fig.~\ref{vcomp:fig}c-d, respectively. The response of the curtain produces velocity in the negative $x$ direction due to reflection and the restoring Lorentz force of the perturbed dipole field (see, Fig.~\ref{vcomp:fig}a) with smaller magnitude than the pulse. The Gaussian pulse launched at the boundary causes rarefaction of the adjacent density region and propagation of nonlinear (steepened) compressive fast magnetosonic wave that perturbes nonlinearly the magnetic structure of the curtain inducing oscillations. The propagation and the reflection of the wave is nonuniform due to the bipolar structure of the magnetic field and the corresponding structure of the fast magnetosonic speed (shown in Figure~\ref{3Dinit:fig}b). As the fast magnetosonic wave leaves the computational domain the oscillations of the curtain proceed to relax approaching normal mode, and eventually damp as the transverse waves leak out of the structure, and the final density and magnetic field structure returns close to the initial structure. 

\begin{figure}
	\centering
   	{\includegraphics[width=9cm]{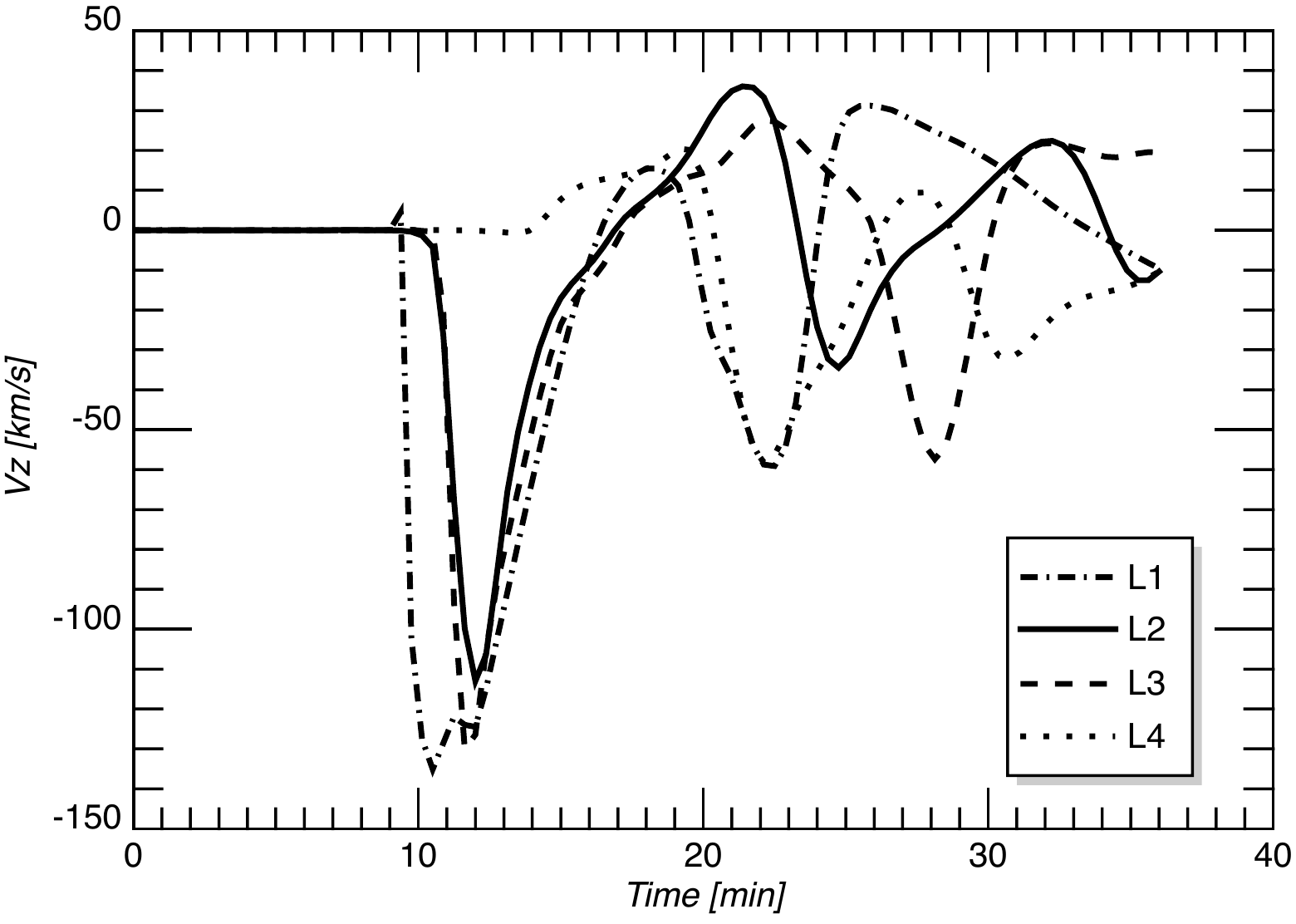}}
 	\caption{Temporal evolution of the model velocity fluctuations in the \textit{yz} plane at different locations along the \textit{x} axis. L1: near flare blast deep layer oscillations $(-1.75,-3.35,1.15)$; L2: near apex deep layer oscillations $(0.0,1.7,1.3)$; L3: near apex surface oscillations $(0.0,1.7,1.85)$; L4: southward surface oscillations $(1.75,-0.9,1.65)$. }
 	\label{vtime_model:fig}
\end{figure}

In Fig.~\ref{vtime_model:fig}, we display the temporal evolution of vertical fluctuations in the vertical plasma velocity component $V_z$ at locations $L1-L4$, very similar to what has been shown by observations (Fig.~\ref{SG13:fig}). We observe variations in the wave phase speed at different points, not necessarily aligned with the \textit{y} axis. The reason of this choice of `observing points' in the model comes from direct observations of the event by SG13, suggesting that the bundle of loops experiencing the vertical kink oscillations might be quite tilted to the bottom boundary surface, and the oscillations are not necessarily co-planar. 

The dash-dotted curve in Fig.~\ref{vtime_model:fig} shows vertical transverse kink oscillations near the boundary plane where the flare produced pulse is modeled, i.e., nearest to the flaring site (L1) and taking place in the deeper layers of the plasma curtain ($z=1.3$, $\sim 20$ Mm above the photosphere). This oscillation has been interpreted by SG13 as the first overtone with a period of 8.9 min (cf., Fig.~\ref{SG13:fig}c for velocity time dependence). For oscillations in the northward side of the plasma curtain, our MHD model yields a fundamental period $P_0=2\pi/\omega$ = 9.2 min with 99.99\% confidence level, in surprisingly good agreement with SDO/AIA observations, given the simplicity of the present magnetic field model. 

The solid curve in Fig.~\ref{vtime_model:fig} shows the plasma vertical velocity versus time at locations L2 and the dashed curved shows L3: near the loop apex and mid-way from the flaring site ($y=0$), respectively. In this region there is observational evidence for oscillations in both deep and surface layers of the coronal curtain ($\sim 20$ Mm and $\sim 60$ Mm over the surface). According to SG13, the large-scale disturbance at depth could represent the fundamental kink mode (but, possibly dominated  by the first harmonic, see below) with a period of 13.3 min, while the non-decaying surface disturbance perturbs only the magnetic upper sheet of the curtain and is characterized by a period of 14.9 min. Once again our MHD model of the event seems to reproduce the amplitude and period of the perturbations with a very good agreement: $A$=34.8 km/s, $P_0$ = 13.5 min for the deep mode; $A$=40.1 km/s, $P_0$ = 14.8 min for the surface mode, where the periods are obtained from the periodorgram analysis \citep{Sca82}, and the amplitude is obtained from sinusoidal least-square-fit with close valued of the periods. The time-profile of associated observed velocity perturbations can be seen in Fig.~\ref{SG13:fig}b.
\begin{figure}
	\centering
   	{\includegraphics[width=9cm]{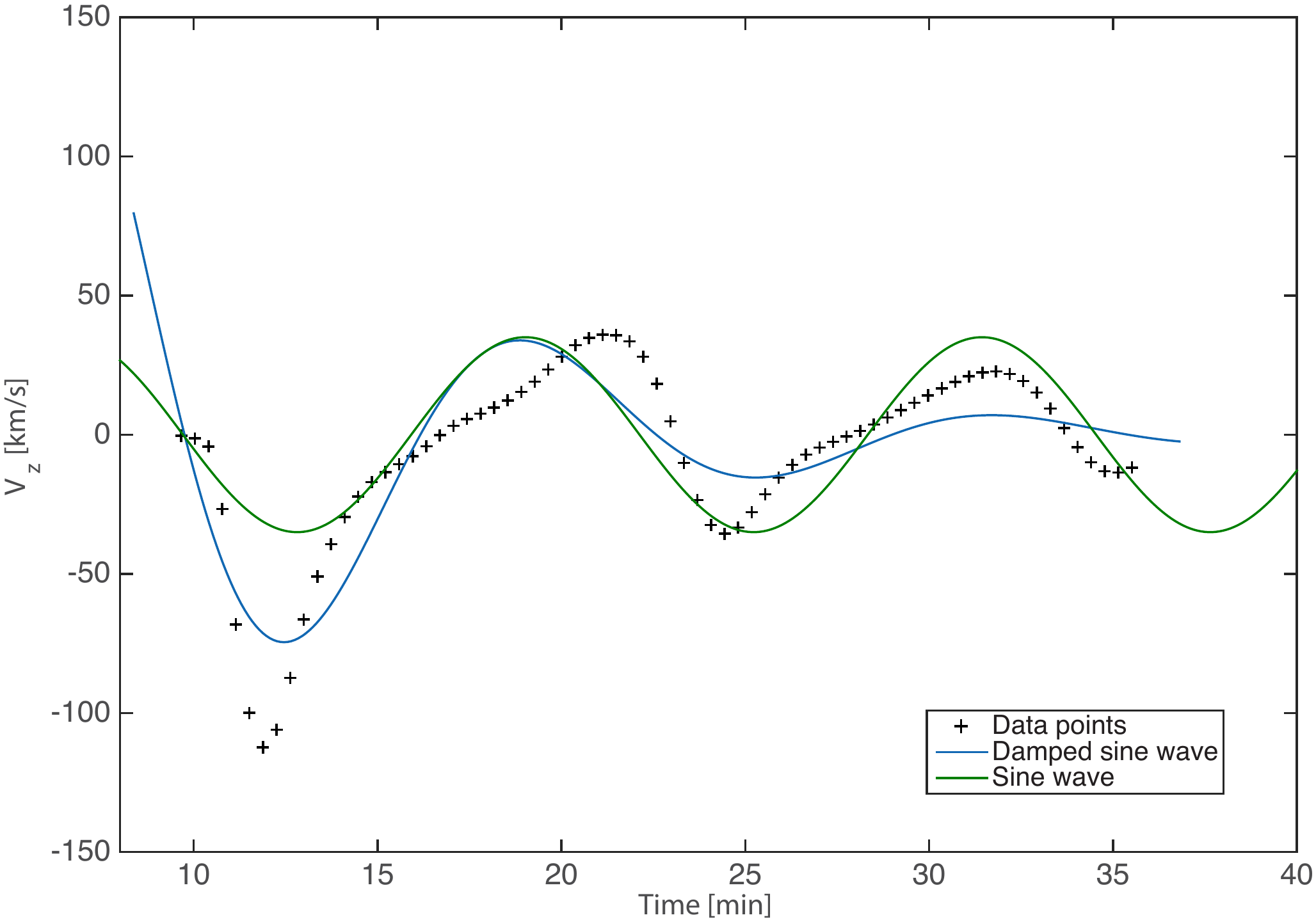}}
 	\caption{Temporal evolution of the model velocity oscillation in the \textit{yz} plane at L2 ('+' symbols) fitted with a single frequency sinusoidal function (green) and the damped sinusoidal function (blue).  The resulting period for the damped sinusoidal is 12.8$\pm0.8$ min and the best-fit damping time is in the range 6.2-11.7 min. The best-fit period is $12.4\pm0.8$ min for the sinusoidal fit.}
 	\label{dsine_fit:fig}
\end{figure}

\begin{figure}
	\centering
   	{\includegraphics[width=9cm]{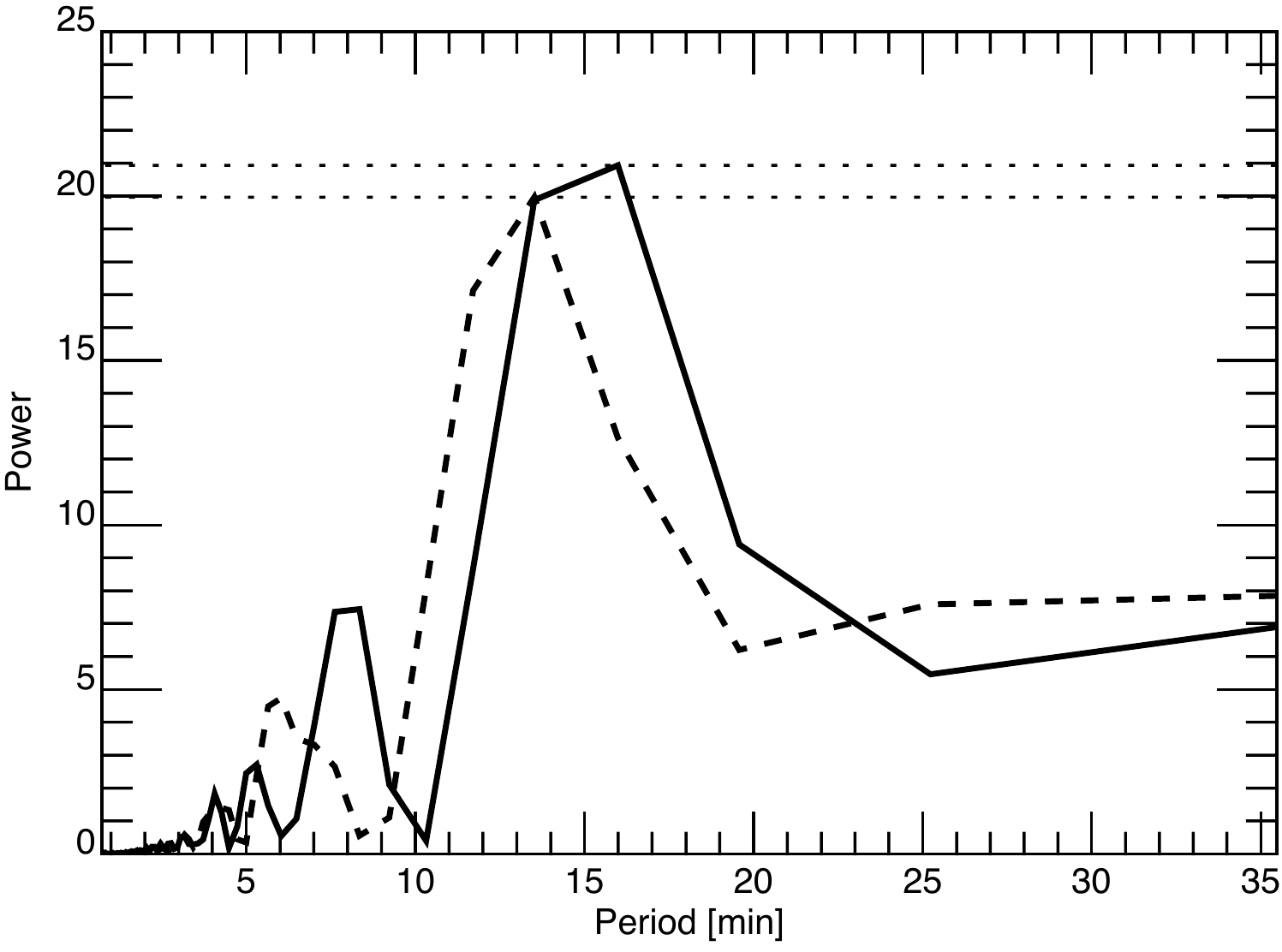}}
 	\caption{The results of the periodogram analysis \citep{Sca82} of the oscillations at L2 (solid) and L3 (dashes). The confidence level of the dominant period is shown with the dash line for each periodogram.}
 	\label{period:fig}
\end{figure}

Lastly, we show perturbations in the top layers of the coronal curtain, L4, shown with short dashes in  Fig.~\ref{vtime_model:fig}, southward and opposite to the flaring site ($y=1.75$). Vertical kink modes have been observed in this region as well, with a period of 12.7 min by SG13. Analysis of   our modeled oscillations at this location yields $A$=46.1 km s$^{-1}$ and $P_0$ = 11.7 min, in good agreement with observational measurements (cf., Fig.~\ref{SG13:fig}b).   The detection of the fundamental and second harmonic in the oscillations is important since their ratio in a coronal loop can be used for the determination of the gravitational scale height of the density \citep[e.g.,][]{And05,Rud08,Ver08}. This effect is expected to be significant in long stratified coronal loop structures. However, in the present study we deal with oscillations of a bipolar structure in short magnetic field line loops (see, Fig.~\ref{3Dinit:fig}), therefore, the magnitude of the second harmonic is found to be weak. Guided by SG13 observations, high-contrast density structure of individual loops is not included in the model of the curtain. 

In Figure \ref{dsine_fit:fig} we show the results of least-square exponentially damped single-frequency sinusoidal fit to the modeled oscillation at L2. The fitted function is of the form
\begin{eqnarray}
v(t)=v_0 e^{-t/\lambda} sin(\omega t+\phi),
\end{eqnarray} where $\lambda$ is the damping time and $\omega=2\pi/P$ is the oscillation frequency. We also show an undamped single-frequency sinusoidal fit. The damped sinusoidal fits produces  a period of 12.8$\pm$0.8 min and the damping time in the range 6.2-11.7 min. The fit produces an amplitude of 355 km s$^{-1}$ and the phase $\phi=4.6$ Rad at $t=0$. However, the values of the amplitude and the phase at $t=0$ are not directly relevant in this analysis, since the oscillations at L2 start at $t\sim9$ min due to the travel time of the pulse from the excitation boundary to L2. Due to the nonlinearity in the initial stages of the oscillations that produces large initial velocity fluctuation, the fitted exponential damping time is overestimated. However, it is interesting to note that the damping time is of the same order as the period of the oscillations. Since the large amplitude of the initial nonlinear pulse may affect the damped sinusoidal fit, we have also fitted undamped sinusoidal to the modeled velocity oscillations. At point L2 we get a period of 12.4$\pm$0.8 min for the undamped sinusoidal. We note that the values of the periods obtained from the damped sinusoidal, undamped sinusoidal, and the periodogram analysis are close within the respective error bars.

The periodogram analysis \citep{Sca82} of the modeled oscillation time sequences for points L2 and L3 are shown in Fig.~\ref{period:fig}. It is evident that the oscillations are dominated by a single frequency with similar results at points L1 and L4 (not shown). While there is some evidence of second and higher harmonics in the analysis, the statistical confidence level of these harmonics is low $<1\sigma$. The low statistical significance of the second harmonic may explain why the second harmonic is rarely detected in observations of loop oscillations. We have also performed curve fitting of single-period sinusoidal and exponentially damped sinusoidal function to the oscillations and found that the values of the fitted periods are in excellent agreement with the periodogram values. We find that the damping time of the oscillations is on the order of the oscillation period.

Table \ref{Table 1} summarizes our results from the model velocity oscillation, and compares them to SG13 results obtained using SDO/AIA data. The modeled error bar of the periods obtained from least-square fitting is below 3\%. However, since there is no statistical variability in a computational model, the meaning of error bars is ill-defined in this context. Therefore, the computational values are shown without error bars. The last two columns of Table~1 show the comparison between the mean magnetic field along the magnetic field lines (loops) that pass through the measurement points in the 3D MHD model ($B_{MHD}$), and the magnetic field inferred from application of coronal seismology (CS), $B_{CS}$. Since the gravitationally stratified density and the magnetic field  vary significantly along the loops \citep[e.g.,][]{OWD12} we use the average values along the magnetic loops. We are comparing the values of the average magnetic field along the magnetic loops that pass through the points L1-L4 in our MHD model, with the average value of the magnetic field determined from coronal seismology that uses a slab model for these loops (Equation~\ref{bcs:eq}). Since the density along a loop is close to the surrounding density we use  $C_k\approx V_A$ in the low-$\beta$ plasma, and we calculate the loop-averaged density to find $B_{CS}$. Since the wavelength of the fundamental mode oscillations is twice the loop length, the waves cannot be used to determine magnetic field value at a particular point of the variable magnetic field in the loop (see, Figure 3 in \citet{OWD12},  for an example of magnetic field, density, and fast magnetosonic speed variability along a loop in the bipolar field model). The long wavelength oscillations provide information only on the average magnetic field in such loops, and these value are given in Table~1.

The event animation from SDO/AIA shows how the plasma curtain oscillations are combinations of the fundamental and second normal modes. However, the latter contribution seems to be predominant at deep and surface levels. Therefore, CS predictions are computed using the expression for second harmonic oscillations of the ideal vertical kink mode without density contrast across the loop (consistent with observations that show little density contrast), and average values along the loops \vspace{-0.01cm}\begin{eqnarray}
	B_{CS} = \left<\rho\right>^{1/2} \frac{L}{P_0},
	\label{bcs:eq}
\end{eqnarray}
where $\left<\rho\right>$ is the average density of the loop, $L$ is the loop length,  $P_0$ is the oscillation period in the MHD model, and $B_{CS}$ is the average magnetic field magnitude along the loop.

\begin{table*}
	\centering
	\begin{tabular}{l c c c c}
	\hline
		Location of observation & Observed periods [min] & MHD period [min] & $B_{MHD}$ (G) & $B_{CS}$ (G)\\
		\hline
		Flare Blast (L1) & 8.9 & 9.2 & 2.2 & 8.4\\
		Apex Inside (L2) & 13.3 & \ 13.5$^*$ & 3.1 & 4.3\\
		Apex Surface (L3) & 14.9 & 14.8 & 2.9 & 4.3\\
		Southward Surface (L4) & 12.7 & 11.7 & 3.4 & 4.9\\
		\hline
\end{tabular}
\caption{Comparison between observed and model oscillation periods for the four detected location in SG13.  The values of the MHD periods were obtained using periodogram analysis \citep{Sca82} and the period with the highest intensity was chosen as the proxy for the normal mode.  The penultimate  column shows the magnetic field strength obtained from the 3D MHD model, and the last column is from simplified coronal seismology (CS). $^*$At point L2 the average of two periods near the peak were taken since they have close high confidence level (see, Fig.~\ref{period:fig}).}
\label{Table 1}
\end{table*}


The periods of the observed and modeled oscillations show good agreement. Comparing the MHD and coronal seismology predictions of the mean magnetic field, we  conclude that the differences are within 39\% - 47\% at 3 out of 4 observation locations (Fig.~\ref{vtime_model:fig}b-d and row 2-4 in Table~\ref{Table 1}). This level of agreement was also found by \citet{AS11} when comparing $B_{CS}$ values to extrapolated magnetogram field values and by \citet{PD14} in 3D MHD arcade model test of CS. The poorer agreement between the  values near the Flare Blast location (Fig.~\ref{vtime_model:fig}a and row 1 in Table \ref{Table 1}, $B_{CS} / B_{MHD} \approx 4$) is likely due to two factors: the large tilt of the magnetic loop that passes  through the observation point to the \textit{yz} plane, causing the loop to interact with the lower boundary surface, and the possibility that the observed oscillation is not a normal harmonic but rather a driven propagating mode due to the close proximity of the boundary that launches the Gaussian velocity pulse.

\section{Discussion and Conclusions}
\label{con:sec}
The article demonstrates that an ideal dipole approximation of the magnetic field geometry is a good approximation of the magnetic field structure in an AR plasma curtain on a large spatial scale over the western equatorial corona observed on 2011 August 9 by SDO/AIA. This large-scale magneto-plasma system, when encountered with the flare generated propagating disturbances, produces multiple harmonics of vertical kink oscillations in the different parts of the curtain, as well as the 
surface oscillations. The 3D MHD numerical simulation domain  includes a magneto-plasma system  with gravitationally stratified density, and its interaction with the velocity pulse introduced at the boundary mimicking the flare disturbance, which propagates into the modeled AR curtain and produces magnetic field displacement and velocity oscillations of vertical kink waves. The results of the numerical simulation match well  with the observations of SG13, while it is initialized with parameters based on the observed conditions, and thus improve the understanding of the wave activity. 

While the present 3D MHD model produces overall evolution of the vertical transverse waves in good agreement with observations - the scope of the model is limited, and is not intended to study flares and the detailed heating and cooling dynamics of an active regions. The main limitation of the model is the use of dipole field that is only approximately applicable to this observation, and this could be improved by using extrapolated magnetogram field \citep[e.g.][]{Ofm07} in future studies. The detailed effects of coronal heating, thermal conduction, and radiative cooling are not modeled and only approximately accounted. The use of polytropic energy equation with $\gamma=1$ is only applicable to approximately isothermal conditions where the above effects nearly balance and the heat conduction is high along the field. The initially hydrostatic density structure while providing overall reasonable approximation, does not capture the full complexity of active region and individual loop density structure and neglects the potentially important role of upflows and downflows that can affect the wave dynamics.

The damping of the oscillations in the model is due to refraction and leakage of the fast mode wave flux as demonstrated in previous work \citep{OT02,MO08,SO09,SOS11} since non-ideal damping processes, such as resistive or numerical dissipation are small in the present model.  It is evident that the observed oscillations  appear to damp on a timescale similar to the modeled evolution. The effects of mode conversion (i.e., resonant absorption) as discussed by SG13 are possible but not strongly supported since the observed AR curtain shows little evidence of individual loop structures (with high density contrast) necessary for this damping mechanism. The modeled low-$\beta$ plasma curtain contains strong magnetic field and the density is gravitationally stratified to mimic the observations and to best fit 
with the realistic solar atmosphere. Therefore, a significant fast-mode speed gradient is set across the plasma curtain from its deeper layers towards the apex. This may set in a possibility of internal refraction of the waves and transfer of the fast mode (kink) wave energy towards the surface and outside the curtain \citep[as demonstrated numerically by, e.g.,][]{OT02}. Somewhat alternative interpretation of the refraction process can be found in \citet{Goss02,Goss11,Goss12}, adapted by SG13 initial analysis.

The 3D MHD model of the waves in the coronal curtain allows us to test the application of simplified coronal seismology as it is widely used \citep[see the reviews, e.g.,][]{NV05,LO14} to the present model and compare the results of CS with the actual values of the magnetic field used in the 3D MHD model. We find that simplified coronal seismology provides magnetic field values within 40\%-50\% of the actual values. Noting,  that the magnetic energy of AR important for Space Weather impact analysis scales as $B^2$, this discrepancy amount to factor of $\sim$2 of the magnetic energy content. This result is consistent with previous tests  of CS in more simplified setup \citep[e.g.,][in magnetic arcade without gravity]{PD14} and with comparisons to extrapolated magnetogram field \citep{AS11}. Thus, in order to improve CS and provide more accurate values of the magnetic field in ARs it is necessary to consider the properties of waves in complex, curved, and more realistic magnetic structures with realistic global density structure as demonstrated in the present study.

\begin{acknowledgements} We are grateful to SDO/AIA team for providing the
data used in this study. LO was supported by NASA grants NNX11AO68G and NNX12AB34G.
AKS acknowledge the patient encouragements of Shobhna. We thank T.J. Wang for useful discussions.
\end{acknowledgements}


\begin{thebibliography}{55}
\expandafter\ifx\csname natexlab\endcsname\relax\def\natexlab#1{#1}\fi

\bibitem[{{Al-Ghafri} {et~al.}(2014){Al-Ghafri}, {Ruderman}, {Williamson}, \&
  {Erd{\'e}lyi}}]{Alg14} 
{Al-Ghafri}, K.~S., {Ruderman}, M.~S., {Williamson}, A., \& {Erd{\'e}lyi}, R.
  2014, \apj, 786, 36

\bibitem[{{Andries} {et~al.}(2005){Andries}, {Goossens}, {Hollweg}, {Arregui},
  \& {Van Doorsselaere}}]{And05}
{Andries}, J., {Goossens}, M., {Hollweg}, J.~V., {Arregui}, I., \& {Van
  Doorsselaere}, T. 2005, \aap, 430, 1109

\bibitem[{{Andries} {et~al.}(2009){Andries}, {van Doorsselaere}, {Roberts},
  {Verth}, {Verwichte}, \& {Erd{\'e}lyi}}]{And09}
{Andries}, J., {van Doorsselaere}, T., {Roberts}, B., {et~al.} 2009, \ssr, 149,
  3

\bibitem[{{Aschwanden}(2005)}]{Asc05}
{Aschwanden}, M.~J. 2005, {Physics of the Solar Corona. An Introduction with
  Problems and Solutions (2nd edition)} (Springer-Praxis)

\bibitem[{{Aschwanden} \& {Boerner}(2011)}]{AscB11}
{Aschwanden}, M.~J. \& {Boerner}, P. 2011, \apj, 732, 81

\bibitem[{{Aschwanden} {et~al.}(1999){Aschwanden}, {Fletcher}, {Schrijver}, \&
  {Alexander}}]{Asc99}
{Aschwanden}, M.~J., {Fletcher}, L., {Schrijver}, C.~J., \& {Alexander}, D.
  1999, \apj, 520, 880

\bibitem[{{Aschwanden} \& {Schrijver}(2011)}]{AS11}
{Aschwanden}, M.~J. \& {Schrijver}, C.~J. 2011, \apj, 736, 102

\bibitem[{{De Moortel} \& {Nakariakov}(2012)}]{DN12}
{De Moortel}, I. \& {Nakariakov}, V.~M. 2012, Royal Society of London
  Philosophical Transactions Series A, 370, 3193

\bibitem[{{Erd{\'e}lyi} {et~al.}(2011){Erd{\'e}lyi}, {Al-Ghafri}, \&
  {Morton}}]{Erd11}
{Erd{\'e}lyi}, R., {Al-Ghafri}, K.~S., \& {Morton}, R.~J. 2011, \solphys, 272,
  73

\bibitem[{{Goossens} {et~al.}(2002){Goossens}, {Andries}, \&
  {Aschwanden}}]{Goss02}
{Goossens}, M., {Andries}, J., \& {Aschwanden}, M.~J. 2002, \aap, 394, L39

\bibitem[{{Goossens} {et~al.}(2012){Goossens}, {Andries}, {Soler}, {Van
  Doorsselaere}, {Arregui}, \& {Terradas}}]{Goss12}
{Goossens}, M., {Andries}, J., {Soler}, R., {et~al.} 2012, \apj, 753, 111

\bibitem[{{Goossens} {et~al.}(2011){Goossens}, {Erd{\'e}lyi}, \&
  {Ruderman}}]{Goss11}
{Goossens}, M., {Erd{\'e}lyi}, R., \& {Ruderman}, M.~S. 2011, \ssr, 158, 289

\bibitem[{{Gruszecki} {et~al.}(2006){Gruszecki}, {Murawski}, {Selwa}, \&
  {Ofman}}]{Gru06}
{Gruszecki}, M., {Murawski}, K., {Selwa}, M., \& {Ofman}, L. 2006, \aap, 460,
  887

\bibitem[{{Kim} {et~al.}(2014){Kim}, {Nakariakov}, \& {Cho}}]{Kim14}
{Kim}, S., {Nakariakov}, V.~M., \& {Cho}, K.-S. 2014, \apjl, 797, L22

\bibitem[{{Lemen} {et~al.}(2012){Lemen}, {Title}, {Akin}, {Boerner}, {Chou},
  {Drake}, {Duncan}, {Edwards}, {Friedlaender}, {Heyman}, {Hurlburt}, {Katz},
  {Kushner}, {Levay}, {Lindgren}, {Mathur}, {McFeaters}, {Mitchell}, {Rehse},
  {Schrijver}, {Springer}, {Stern}, {Tarbell}, {Wuelser}, {Wolfson}, {Yanari},
  {Bookbinder}, {Cheimets}, {Caldwell}, {Deluca}, {Gates}, {Golub}, {Park},
  {Podgorski}, {Bush}, {Scherrer}, {Gummin}, {Smith}, {Auker}, {Jerram},
  {Pool}, {Soufli}, {Windt}, {Beardsley}, {Clapp}, {Lang}, \&
  {Waltham}}]{Lem12}
{Lemen}, J.~R., {Title}, A.~M., {Akin}, D.~J., {et~al.} 2012, \solphys, 275, 17

\bibitem[{{Liu} \& {Ofman}(2014)}]{LO14}
{Liu}, W. \& {Ofman}, L. 2014, \solphys, 289, 3233

\bibitem[{{McIntosh} {et~al.}(2011){McIntosh}, {de Pontieu}, {Carlsson},
  {Hansteen}, {Boerner}, \& {Goossens}}]{Mci11}
{McIntosh}, S.~W., {de Pontieu}, B., {Carlsson}, M., {et~al.} 2011, \nat, 475,
  477

\bibitem[{{McLaughlin} \& {Ofman}(2008)}]{MO08}
{McLaughlin}, J.~A. \& {Ofman}, L. 2008, \apj, 682, 1338

\bibitem[{{Miyagoshi} {et~al.}(2004){Miyagoshi}, {Yokoyama}, \&
  {Shimojo}}]{Miy04}
{Miyagoshi}, T., {Yokoyama}, T., \& {Shimojo}, M. 2004, \pasj, 56, 207

\bibitem[{{Morton} \& {Erd{\'e}lyi}(2009)}]{ME09}
{Morton}, R.~J. \& {Erd{\'e}lyi}, R. 2009, \apj, 707, 750

\bibitem[{{Nakariakov} {et~al.}(1999){Nakariakov}, {Ofman}, {DeLuca},
  {Roberts}, \& {Davila}}]{Nak99}
{Nakariakov}, V.~M., {Ofman}, L., {DeLuca}, E., {Roberts}, B., \& {Davila},
  J.~M. 1999, Science, 285, 862

\bibitem[{{Nakariakov} \& {Verwichte}(2005)}]{NV05}
{Nakariakov}, V.~M. \& {Verwichte}, E. 2005, Living Reviews in Solar Physics,
  2, 3

\bibitem[{{Nistic{\`o}} {et~al.}(2014){Nistic{\`o}}, {Anfinogentov}, \&
  {Nakariakov}}]{Nis14}
{Nistic{\`o}}, G., {Anfinogentov}, S., \& {Nakariakov}, V.~M. 2014, \aap, 570,
  A84

\bibitem[{{Nistic{\`o}} {et~al.}(2013){Nistic{\`o}}, {Nakariakov}, \&
  {Verwichte}}]{Nis13}
{Nistic{\`o}}, G., {Nakariakov}, V.~M., \& {Verwichte}, E. 2013, \aap, 552, A57

\bibitem[{{Ofman}(2007)}]{Ofm07}
{Ofman}, L. 2007, \apj, 655, 1134

\bibitem[{{Ofman}(2009)}]{Ofm09}
{Ofman}, L. 2009, \apj, 694, 502

\bibitem[{{Ofman} \& {Thompson}(2002)}]{OT02}
{Ofman}, L. \& {Thompson}, B.~J. 2002, \apj, 574, 440

\bibitem[{{Ofman} {et~al.}(2012{\natexlab{a}}){Ofman}, {Wang}, \&
  {Davila}}]{OWD12}
{Ofman}, L., {Wang}, T.~J., \& {Davila}, J.~M. 2012{\natexlab{a}}, \apj, 754,
  111

\bibitem[{{Pascoe} \& {De Moortel}(2014)}]{PD14}
{Pascoe}, D.~J. \& {De Moortel}, I. 2014, \apj, 784, 101

\bibitem[{{Powell}(1994)}]{Pow94}
{Powell}, K.~G. 1994, Report NM-R9407

\bibitem[{{Ruderman}(2011)}]{Rud11}
{Ruderman}, M.~S. 2011, \aap, 534, A78

\bibitem[{{Ruderman} {et~al.}(2008){Ruderman}, {Verth}, \&
  {Erd{\'e}lyi}}]{Rud08}
{Ruderman}, M.~S., {Verth}, G., \& {Erd{\'e}lyi}, R. 2008, \apj, 686, 694

\bibitem[{{Scargle}(1982)}]{Sca82}
{Scargle}, J.~D. 1982, \apj, 263, 835

\bibitem[{{Schmidt} \& {Ofman}(2010)}]{SO10}
{Schmidt}, J.~M. \& {Ofman}, L. 2010, \apj, 713, 1008

\bibitem[{{Selwa} {et~al.}(2007){Selwa}, {Murawski}, {Solanki}, \&
  {Wang}}]{Sel07}
{Selwa}, M., {Murawski}, K., {Solanki}, S.~K., \& {Wang}, T.~J. 2007, \aap,
  462, 1127

\bibitem[{{Selwa} {et~al.}(2005){Selwa}, {Murawski}, {Solanki}, {Wang}, \&
  {T{\'o}th}}]{Sel05}
{Selwa}, M., {Murawski}, K., {Solanki}, S.~K., {Wang}, T.~J., \& {T{\'o}th}, G.
  2005, \aap, 440, 385

\bibitem[{{Selwa} \& {Ofman}(2009)}]{SO09}
{Selwa}, M. \& {Ofman}, L. 2009, Annales Geophysicae, 27, 3899

\bibitem[{{Selwa} {et~al.}(2011{\natexlab{a}}){Selwa}, {Ofman}, \&
  {Solanki}}]{SOS11}
{Selwa}, M., {Ofman}, L., \& {Solanki}, S.~K. 2011{\natexlab{a}}, \apj, 726, 42

\bibitem[{{Selwa} {et~al.}(2013){Selwa}, {Poedts}, \& {DeVore}}]{Sel13}
{Selwa}, M., {Poedts}, S., \& {DeVore}, C.~R. 2013, \solphys, 284, 515

\bibitem[{{Selwa} {et~al.}(2006){Selwa}, {Solanki}, {Murawski}, {Wang}, \&
  {Shumlak}}]{Sel06}
{Selwa}, M., {Solanki}, S.~K., {Murawski}, K., {Wang}, T.~J., \& {Shumlak}, U.
  2006, \aap, 454, 653

\bibitem[{{Selwa} {et~al.}(2011{\natexlab{b}}){Selwa}, {Solanki}, \&
  {Ofman}}]{SSO11} {Selwa}, M., {Solanki}, S.~K., \& {Ofman}, L. 2011{\natexlab{b}}, \apj, 728, 87

\bibitem[{{Srivastava} \& {Goossens}(2013)}]{SG13}
{Srivastava}, A.~K. \& {Goossens}, M. 2013, \apj, 777, 17

\bibitem[{{Van Doorsselaere} {et~al.}(2011){Van Doorsselaere}, {Wardle}, {Del
  Zanna}, {Jansari}, {Verwichte}, \& {Nakariakov}}]{Van11}
{Van Doorsselaere}, T., {Wardle}, N., {Del Zanna}, G., {et~al.} 2011, \apjl,
  727, L32

\bibitem[{{Verth} {et~al.}(2008){Verth}, {Erd{\'e}lyi}, \& {Jess}}]{Ver08}
{Verth}, G., {Erd{\'e}lyi}, R., \& {Jess}, D.~B. 2008, \apjl, 687, L45

\bibitem[{{Verwichte} {et~al.}(2006){Verwichte}, {Foullon}, \&
  {Nakariakov}}]{Ver06}
{Verwichte}, E., {Foullon}, C., \& {Nakariakov}, V.~M. 2006, \aap, 452, 615

\bibitem[{{Verwichte} {et~al.}(2013{\natexlab{a}}){Verwichte}, {Van
  Doorsselaere}, {Foullon}, \& {White}}]{Ver13b}
{Verwichte}, E., {Van Doorsselaere}, T., {Foullon}, C., \& {White}, R.~S.
  2013{\natexlab{a}}, \apj, 767, 16

\bibitem[{{Verwichte} {et~al.}(2013{\natexlab{b}}){Verwichte}, {Van
  Doorsselaere}, {White}, \& {Antolin}}]{Ver13a}
{Verwichte}, E., {Van Doorsselaere}, T., {White}, R.~S., \& {Antolin}, P.
  2013{\natexlab{b}}, \aap, 552, A138

\bibitem[{{Wang} {et~al.}(2013){Wang}, {Ofman}, \& {Davila}}]{WOD13}
{Wang}, T., {Ofman}, L., \& {Davila}, J.~M. 2013, \apjl, 775, L23

\bibitem[{{Wang} {et~al.}(2012){Wang}, {Ofman}, {Davila}, \& {Su}}]{Wan12}
{Wang}, T., {Ofman}, L., {Davila}, J.~M., \& {Su}, Y. 2012, \apjl, 751, L27

\bibitem[{{Wang} \& {Solanki}(2004)}]{WS04}
{Wang}, T.~J. \& {Solanki}, S.~K. 2004, \aap, 421, L33

\bibitem[{{Wang} {et~al.}(2008){Wang}, {Solanki}, \& {Selwa}}]{WSS08}
{Wang}, T.~J., {Solanki}, S.~K., \& {Selwa}, M. 2008, \aap, 489, 1307

\bibitem[{{White} \& {Verwichte}(2012)}]{WV12}
{White}, R.~S. \& {Verwichte}, E. 2012, \aap, 537, A49

\bibitem[{{White} {et~al.}(2012){White}, {Verwichte}, \& {Foullon}}]{WVF12}
{White}, R.~S., {Verwichte}, E., \& {Foullon}, C. 2012, \aap, 545, A129

\bibitem[{{Yuan} \& {Nakariakov}(2012)}]{YN12}
{Yuan}, D. \& {Nakariakov}, V.~M. 2012, \aap, 543, A9

\end{thebibliography}


\end{document}